\documentclass[letterpaper, 10 pt, conference]{ieeeconf}  

\IEEEoverridecommandlockouts                              

\usepackage[utf8]{inputenc}     

\usepackage[T1]{fontenc}        

\usepackage[english]{babel}              

\usepackage{graphicx}           
\usepackage{amsmath,amssymb}    
\usepackage{color}
\usepackage[usenames]{xcolor}   
\usepackage{microtype} 

\usepackage{marginnote}         

\usepackage{mathtools}
\usepackage{siunitx}
\usepackage{pbox}
\usepackage{dsfont}
\usepackage{algorithm}

\usepackage{tikz}				
\usetikzlibrary{arrows}				
\usetikzlibrary{shapes.misc}
\usetikzlibrary{patterns}
\usepackage{pgfplots}
\usetikzlibrary{calc}

\colorlet{istorange}{orange}
\colorlet{istgreen}{green!50!black}
\colorlet{istblue}{blue} 
\colorlet{istred}{red!90!black}

\newtheorem{theorem}{Theorem}
\newtheorem{lemma}{Lemma}

\newtheorem{assumption}{Assumption}
\newtheorem{definition}{Definition}

\newtheorem{remark}{Remark}

\newtheorem{Algorithm}{\bf Algorithm}

\newcommand{\X}{\mathbb{X}}
\newcommand{\U}{\mathbb{U}}

\newcommand{\I}{\mathbb{I}}
\newcommand{\R}{\mathbb{R}}
\newcommand{\W}{\mathbb{W}}
\newcommand{\N}{\mathbb{N}}

\makeatletter
\renewcommand*{\@opargbegintheorem}[3]{\trivlist
	\item[\hskip \labelsep{\textit{ #1\ #2}}] \textit{(#3)}: }
\makeatother

\title{\LARGE \bf
Rollout scheduling and control for disturbed systems via tube MPC: extended version$^*$}

\author{Stefan Wildhagen and Frank Allg{\"o}wer
	\thanks{$^*$This paper is an extended version of \cite{Wildhagen20_3} containing all proofs. Funded by the Deutsche Forschungsgemeinschaft (DFG, German Research Foundation) - 285825138; 390740016. The authors are with the Institute for Systems Theory and Automatic Control, University of Stuttgart, Germany.
		{\tt\small \{wildhagen}{\tt\small,allgower\}@ist.uni}{\tt\small -stuttgart.de}.} %
}

\begin{document}

\maketitle
\thispagestyle{empty}
\pagestyle{empty}

\begin{abstract}
Rollout control is an MPC-based control method, in which a controller is periodically activated to schedule the transmission of sensor or actuator data. Therein, a preassigned traffic specification acts as a constraint on the scheduled transmissions, ensuring that they are triggered at an admissible rate for the underlying communication network. In this paper, we extend the theory of rollout control by considering bounded disturbances on the controlled plant in the presence of state and input constraints. We use methods from tube MPC, where the error between the nominal (undisturbed) system, used as a prediction model, and the real system is kept in a robust control invariant set. This approach requires satisfaction of a \emph{maximum} inter-transmission interval in closed loop, a guarantee that is not straightforward to obtain in rollout control. Our main contribution is to introduce a novel tube MPC scheme adapted for rollout control, which contains an additional constraint on the predicted transmission schedule. We show that by virtue of this schedule constraint, the inter-transmission interval is bounded in closed loop. We also establish recursive feasibility of this novel MPC scheme and show convergence of the system.
\end{abstract}

\section{Introduction}

Cyber-physical systems, in which physical processes are governed by digital devices, experience a continuously increasing prevalence in application and industry. Since the components of cyber-physical systems are in most cases spatially distributed, coordination and
control of the individual components require the communication of sensor measurements and actuation signals over a common medium, i.e., a digital communication network. In contrast to classical hard-wired, analog communication links, such networks may feature unwanted characteristics such as a limited bandwidth and, especially when congested, transmission delay or packet loss. These challenges gave rise to the research direction of Networked Control Systems (NCSs), which explicitly considers the network as an additional component of the control loop, and studies its impact on the satisfaction of control goals, e.g., stability and control performance.

To ensure a reliable Quality of Service (QoS) of the communication network, it is typically crucial that the adjacent control applications do not congest the network with their transmissions, i.e., that they do not demand a higher data rate than the network is able to provide. Otherwise, if the QoS deteriorates, the Quality of Control (QoC) might suffer with it and even become unpredictable. One way to handle this requirement is to assign a certain share of transmission rate to a control application beforehand, which is captured in a so-called traffic specification (TS). Typically, these specifications do not allow a transmission at every sampling instance, but they do allow that  transmissions are scheduled by the applications so as to maximize control performance. In the literature, TSs like in \cite{Antunes14,Gommans17} (referred to herein as window-based TS) and the token bucket TS \cite{Tanenbaum11,Linsenmayer18} were considered recently.

Predictive, optimization-based controllers can integrate such TSs directly as constraints in the optimization. Two model predictive control (MPC)-based approaches which may handle a TS have emerged in the literature: self-triggered MPC \cite{Henriksson15,Brunner14} and rollout control \cite{Antunes14,Gommans17,Koegel19,Wildhagen19_2}. In the former, the controller is only activated at transmission time instances, such that both transmissions from sensor to controller and from controller to actuator fulfill a given TS. The latter, in contrast, assumes that only one of these communication links must fulfill a TS, while the other is unrestricted. In practice, this requires that either the actuator or the sensor has computational abilities and may implement a controller. The controller is then activated periodically or in each time step. If the NCS architecture allows to apply rollout approaches, a reason to prefer them over self-triggered MPC is that since the controller continues to operate in between transmission instances, they may react more quickly to disturbances.

Nonetheless, there is a number of self-triggered MPC approaches that guarantee satisfaction of state- and input constraints and convergence in the presence of bounded disturbances (e.g., \cite{Brunner14,Zhan17}). In the case of rollout approaches, the picture for disturbed systems is much less complete: In \cite{Koegel19}, the authors consider a robust rollout approach in which sensor data must fulfill a TS, while the communication between controller and actuator is unrestricted. In \cite{Antunes14}, as in the present work, a TS between controller and actuator is considered. However, unconstrained systems and stochastic disturbances are regarded therein, such that the derived results on stability and performance provide purely stochastic guarantees.

The above-mentioned works, except \cite{Antunes14}, make use of so-called tube MPC \cite{Mayne05}. Therein, the MPC controls a nominal model of the plant, disregarding the disturbances, while the crucial idea is that the error between real and nominal system is kept in a so-called robust control invariant (RCI) set (a ``tube'') with the help of an additional error feedback term. When using tube MPC in case the communication between controller and actuator is restricted, one must take into account that due to the constraints of the TS, there might be no control update available for several consecutive time steps. Hence, the error must be kept in the RCI set despite the fact that the error feedback is applied in open loop. Such ``multi-step'' RCI sets already appeared in the literature on robust self-triggered MPC for linear systems \cite{Brunner14,Zhan17}, which may handle any inter-transmission interval in between 1 and $H\in\N$ steps. If such a set and feedback are known, all that is left to do is to ensure that in closed loop, the phases in between transmissions are no longer than $H$. Since in self-triggered MPC, the controller is activated at transmission instants and the prediction horizon spans exactly the time until the next triggering instant, this guarantee is obtained simply by a prediction horizon less than or equal to $H$.

In this paper, we consider an NCS setup where the communication between sensor and controller is unrestricted, but new control inputs must be transmitted to the actuators under satisfaction of a TS. The controlled plant is subject to bounded additive disturbances and state and input constraints. We use a rollout approach, where a controller with cyclically shrinking horizon (cf. \cite{Koegel13}) is activated at each time instant, to schedule transmissions and compute the corresponding control updates simultaneously. In such a setup, the controller is activated independently of transmission instances and the length of the prediction horizon is purely time-dependent, such that at times, it might not even span the time until the next transmission. As a result, guaranteeing a certain maximum inter-transmission interval in closed loop is much more challenging than for self-triggered MPC.

Our main contribution is to enable rollout control to handle bounded disturbances in the presence of state and input constaints. On a technical level, we
adapt tube MPC approaches by, first, giving a suitable, more general definition of a $[1,H]$ RCI set for NCS. Second, we introduce a schedule constraint in the prediction and prove that it ensures a maximum inter-transmission interval of $H$ in closed loop. Third, we propose an adapted tube MPC scheme, in which in contrast to \cite{Mayne05}, the nominal system's initial condition is an optimization variable only in case a transmission is scheduled at initial time. Further, we provide an analysis of recursive feasibility of the novel scheme with the additional schedule constraint, and establish convergence using methods of economic tube MPC \cite{Bayer14,Dong18_2}.

The remainder of this paper is organized as follows: In Section \ref{sec_setup}, we present the considered NCS architecture, where the transmissions between controller and actuator need to satisfy a certain TS. In Section \ref{sec_RCI_set}, we give the definition of a $[1,H]$ RCI set for this NCS. We introduce the schedule constraint and the control scheme in Section \ref{sec_MPC_scheme} before we comment on recursive feasibility, robust constraint satisfaction and convergence in Section \ref{sec_rec_feas_conv}. A numerical example is given in Section \ref{sec_num_ex} to demonstrate the derived results before a summary and an outlook are given in Section \ref{sec_summary_outlook}. Throughout the main body of this paper, we focus on the token bucket TS \cite{Tanenbaum11}, but in Section \ref{subsec_antunes_traffic_spec}, we will argue that all derived results hold for the window-based TS used in \cite{Antunes14,Gommans17}, too.

Let $\I$ and $\R$ denote the set of all integers and real numbers, respectively. We denote $\I_{[a,b]}\coloneqq\I\cap[a,b]$ and $\I_{\ge a}\coloneqq\I\cap[a,\infty)$, $a,b\in\I$. We denote by $A>0$ $(A\ge 0)$ a symmetric positive (semi-)definite matrix $A\in\R^{n\times n}$. For a function $f:\R^n\rightarrow\R^n$, define the image of the set $S\subseteq\R^n$ as $f(S)\coloneqq\{f(x):x\in S\}$. A compact and convex set containing the origin is called a $\mathcal{C}$ set. We denote by $\oplus$ the Minkowski set addition and by $\ominus$ the Pontryagin set difference. The symbols $\lor$, $\land$ denote the logical \emph{or}, \emph{and}, respectively.

\section{Setup} \label{sec_setup}

Consider a disturbed, linear, time-invariant plant
\begin{equation}
x_p(k+1) = Ax_p(k)+Bu_p(k)+w_p(k), \; x_p(0)=x_{p,0}, \label{plant}
\end{equation}
where $k\in\I_{\ge 0}$ denotes the discrete time instance, $x_p(k)$ the state, $u_p(k)$ the input and $w_p(k)$ an additive disturbance which cannot be measured. Both the plant state and plant input are subject to the constraints $x_p(k)\in\X_p\subseteq\R^{n_p}$ and $u_p(k)\in\U_p\subseteq\R^{m_p}$ with closed constraint sets $\X_p$ and $\U_p$ containing the origin. The disturbance acting on the plant is assumed to be bounded within a $\mathcal{C}$ set $\W_p\subseteq\R^{n_p}$, i.e.,
\begin{equation*}
w_p(k)\in \W_p, \; \forall k\in\I_{\ge 0}.
\end{equation*}
Furthermore, a quadratic cost
\begin{equation}
x_p^\top Q x_p + u_p^\top R u_p, \;Q,R>0 \label{cost_plant}
\end{equation}
is associated with the plant as an index for performance.

\begin{figure}
	\centering
	\begin{tikzpicture}[>= latex]
	\node [thick,draw,rectangle, inner sep=1.5pt,minimum size=1.1cm,align = center,rounded corners=3pt] (actuator) at (0.2*7,0) {ZOH Actuator};
	\node[anchor=south east,inner sep=1pt] at (actuator.south east) {\textcolor{istblue}{$u_s$}};
	\node [thick,draw,rectangle, inner sep=1.5pt,minimum size=1.1cm,align = center,rounded corners=3pt] (plant) at (0.55*7,0) {Plant};
	\node[anchor=south east,inner sep=1pt] at (plant.south east) {\textcolor{istblue}{$x_p$}};
	\node [thick,draw,rectangle, inner sep=1.5pt,minimum size = 1.1cm,align = center,rounded corners=3pt] (ctrl) at (0.83*7,0) {MPC};
	\node [thin,dashed,draw,rectangle,minimum height = 1.6cm, minimum width = 2.5cm,align = left] (smartsensor) at (0.895*7,0.07*-2) {\hspace{0.0cm}\\[0.95cm] \hspace{-0.7cm}\small{Smart sensor}};
	\node [thin,dashed,draw,rectangle,minimum height = 1.5cm, minimum width = 3.5cm,align = left] (smartsensor) at (0.5*7,-1.85) {\hspace{0.0cm}\\[1cm] \hspace{-1.65cm}\small{Network}};
	\node [thick,draw,rectangle, inner sep=1.5pt,minimum size = 1.1cm,align = center,rounded corners=3pt] (netw) at (0.5*7,-1.7) {Token Bucket TS};
	\node[anchor=south east,inner sep=1pt] at (netw.south east) {\textcolor{istblue}{$\beta$}};
	
	\node (upper) at (7,0.2*-2){};
	\node (lower) at (7,0.2*-2-0.25){};
	\node (lower2) at (7+0.25*0.5,0.2*-2 -0.25*0.866){};
	\node (dist) at (0.5*7,0.8) {\textcolor{istred}{$w_p$}};
	
	\path (0,0) --++ (7,0) node(helpne){} --++ (0,-1.7) node(helpse){} --++ (-7,0) node(helpsw){} --++ (0,1.7) node(helpnw){};
	
	\draw [->,thick] (plant.east) -- (ctrl.west) node[pos = 0.5,above] {$x_p$};
	\draw [thick] (ctrl.east) -- (helpne.center)node[pos=1.2,above] {$u_c$};
	\draw [thick] (helpne.center) -- (upper.center) -- (lower2.center) node[pos=0.5,right] {$\gamma$};
	\draw[thick] (lower.center) -- (helpse.center) -- (netw.east);
	\draw[->,thick] (netw.west) -- (helpsw.center) -- (helpnw.center) -- (actuator.west);
	\draw[->,thick] (actuator.east) -- (plant.west) node[pos = 0.5,above] {$u_p$};
	
	\draw[->,thick,color=istred] (0.55*7,1) -- (plant.north);
	\end{tikzpicture}
	\caption{Considered configuration of the NCS with token bucket network.}
	\label{fig_network_tb}
\end{figure}
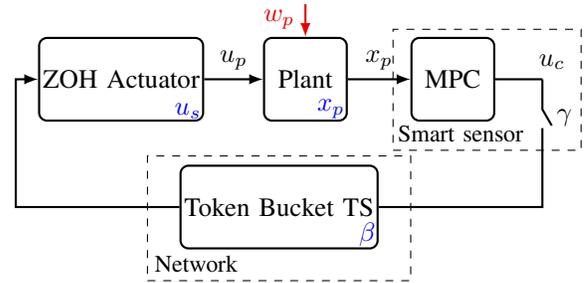
As depicted in Figure \ref{fig_network_tb}, in the NCS setup considered here, an update of the control input $u_c$ cannot be applied to the plant directly, but instead must be transmitted to the actuator via a limited-bandwidth network. Throughout the main body of this paper, we will assume that transmissions over the network must fulfill the so-called token bucket TS \cite{Tanenbaum11}. Therein, tokens are added to a bucket at a rate of $g\in\I_{\ge 1}$ whenever the maximum level of tokens $b\in\I_{\ge c-g}$ is not reached. Otherwise, excessive tokens are discarded. A transmission incurs the cost of $c\in\I_{\ge g}$ tokens (where typically $c>g$, such that a transmission is not possible in every time step). The decision whether to transmit over the network at a given time is taken by the controller to be either $\gamma=1$ (if a transmission is triggered) or $\gamma=0$ (if not). Hence, the bucket level $\beta$ is governed by the difference equation
\begin{equation}
\beta(k+1)=\min\{\beta(k)+g-\gamma(k) c,b\}, \;\beta(0)=\beta_0. \label{token_bucket}
\end{equation}
The token bucket TS is fulfilled if a transmission is only triggered when the bucket level is high enough to support the cost of a transmission, as captured in the constraint $\beta(k)\in\I_{[0,b]}$ for all $k\in\I_{\ge 0}$. We presume throughout this paper that if this constraint is met, transmission delays and packet loss probabilities are negligible. Note that effectively, the token bucket TS prescribes an average transmission rate of $\frac{g}{c}$. 

According to Figure \ref{fig_network_tb}, the state of the disturbed plant $x_p$ is measured by the smart sensor, which runs the MPC and determines both the control input $u_c$ and the transmission decision $\gamma$. Control inputs are sent via the network to a zero-order hold (ZOH) actuator, which holds the last received input and stores it in a state $u_s$ according to
\begin{equation*}
u_s(k+1)=\gamma(k)u_c(k) + (1-\gamma(k))u_s(k)= u_p(k)
\end{equation*}
with $u_s(0)=u_{s,0}\in\U_p$. A more detailed description of the NCS setup and the token bucket TS can be found in \cite{Wildhagen19_2}.

We collect all state and input variables in
\begin{equation*}
x\coloneqq[x_p^\top \; u_s^\top \; \beta]^\top \text{ and } u\coloneqq[u_c^\top \gamma]^\top,
\end{equation*}
and write the nonlinear overall dynamics of the NCS as
\begin{align}
x(k+1) &=f(x(k),u(k))+w(k), \label{system} \\
f(x,u) &\coloneqq \begin{bmatrix} Ax_p + B(1-\gamma) u_s + B\gamma u_c \\ (1-\gamma) u_s + \gamma u_c \\ \min\{\beta+g-\gamma c,b\} \end{bmatrix}, \;
w \coloneqq \begin{bmatrix} w_p \\ 0 \\ 0 \end{bmatrix}. \nonumber
\end{align}
The associated constraints can be formulated as $x(k)\in\X\coloneqq\X_p\times\U_p\times\I_{[0,b]}$, $u(k)\in\U\coloneqq\U_p\times\{0,1\}$ and $w(k)\in\W\coloneqq\W_p\times\{0\}\times\{0\}$. The performance index \eqref{cost_plant} becomes
\begin{equation}
\ell(x,u)\coloneqq x_p^\top Q x_p + (1-\gamma)u_s^\top R u_s + \gamma u_c^\top R u_c. \label{stage_cost}
\end{equation}

\section{$[1,H]$ robust control invariant set} \label{sec_RCI_set}

For the predictions in the tube MPC approach, we rely on the definition of a so-called nominal system
\begin{equation}
\bar{x}(k+1) = f(\bar{x}(k),\bar{u}(k)), \; \bar{x}(0)=\bar{x}_0 \label{nom_system}
\end{equation}
which neglects the disturbance. Approaches in the literature \cite{Bayer14,Dong18_2} apply an error feedback $u(k)=\Phi(\bar{u}(k),x(k),\bar{x}(k))$ to the real system in order to bound the error $x(k)-\bar{x}(k)$ between the real and nominal state within a robust control invariant (RCI) set. Thereby, the influence of all possible disturbance realizations can be captured.

However, unlike in the above-mentioned works, in our setup a control update (and hence an updated version of the error feedback) can in general not be sent to the actuator at each sampling time $k$ due to the constraints of the token bucket TS. For this reason, we consider an error feedback $\phi:\U_p\times\X_p\times\X_p\to\U_p$ which is held by the actuator over $H$ consecutive time steps. We say a set is $[1,H]$ RCI if it has the following property: If the error between real and nominal state is contained in the set at $t=0$ and the zero-order-held error feedback is applied over $H$ time steps, the error does not leave the set in any of these $H$ steps. Hence, in order to capture the influence of the disturbance on the error, we require knowledge of a $[1,H]$ RCI set.

Next, we introduce a formal definition. To this end, consider that the control
\begin{equation}
\bar{v}(0)\coloneqq\begin{bmatrix} \bar{v}_c(0) \\ 1 \end{bmatrix}
, \;
\bar{v}(i)\coloneqq\begin{bmatrix} 0 \\ 0 \end{bmatrix}, \; \forall i\in\I_{[1,H-1]} \label{multi_step_control}
\end{equation}
is applied to the nominal system \eqref{nom_system} with $\bar{v}_c(0)\in\U_p$. Then the error feedback applied to the real system \eqref{system} is
\begin{equation}
\begin{aligned}
\Phi_0(\bar{v}(0),x(0),\bar{x}(0))&\coloneqq\begin{bmatrix} \phi(\bar{v}_c(0),x_p(0),\bar{x}_p(0)) \\ 1 \end{bmatrix} \\
\Phi_i(\bar{v}(i),x(i),\bar{x}(i))&\coloneqq\bar{v}(i)=\begin{bmatrix} 0 \\ 0 \end{bmatrix}, \; \forall i\in\I_{[1,H-1]}.
\end{aligned}
 \label{multi_step_error_feedback}
\end{equation}
Let us consider the state sequences of the nominal system under \eqref{multi_step_control} and of the real system under \eqref{multi_step_error_feedback}, i.e., that $\bar{u}(i)=\bar{v}(i)$ and $u(i)=\Phi_i(\bar{v}(i),x(i),\bar{x}(i))$ for all $i\in\I_{[0,H-1]}$:
\begin{equation}
\begin{aligned}
\bar{x}(i+1) &= f(\bar{x}(i),\bar{v}(i)) \\
x(i+1) &= f(x(i),\Phi_i(\bar{v}(i),x(i),\bar{x}(i)))+ w(i).
\end{aligned} \label{traj}
\end{equation}
The error between the nominal and real system is then for all $i\in\I_{[0,H]}$ defined by
\begin{equation*}
e(i)\coloneqq x(i)-\bar{x}(i),
\end{equation*}
where $\bar{x}(i)$ and $x(i)$ evolve according to \eqref{traj}.

\begin{definition}
	A set $\Omega\subseteq\R^n$ is $[1,H]$ RCI for the error if and only if there exists a $\phi:\U_p\times\X_p\times\X_p\to\U_p$ such that for all $x(0),\bar{x}(0)\in\X$ with $e(0)\in\Omega$, all $\bar{v}_c(0)\in\U_p$ and all $\{w(0),\ldots,w(H-1)\}\in\W^H$ it holds that
	\begin{equation*}
	e(i) \in \Omega, \; \forall i\in\I_{[1,H]}.
	\end{equation*}
\end{definition}

\begin{remark}
	We assume here that $\bar{v}_c(i)=0$ for all $i\in\I_{[1,H]}$. Note, however, that since there is no input transmitted to the actuator when $\gamma=0$, this is without loss of generality.
\end{remark}

Let us define $A_i\coloneqq A^i$ and $B_i\coloneqq\sum_{j=0}^{i-1}A^i B$, $i\in\I_{[0,H]}$.

\begin{lemma}
	Suppose there exist an $H\hspace{-1.7pt}\in\hspace{-1.7pt}\I_{\ge 1}$, a $K\in\R^{m_p\times n_p}$ and a set $\Omega_p\subseteq\R^{n_p}$, containing the origin, which satisfy
	\begin{equation}
	(A_i+B_i K)\Omega_p \oplus \bigg(\bigoplus_{j=0}^{i-1} A_j \W_p\bigg) \subseteq \Omega_p, \forall i\in\I_{[1,H]}. \label{inv_ZOH}
	\end{equation}
	Then, with the feedback law $\phi(\bar{v}_c,x_p,\bar{x}_p)=\bar{v}_c + K(x_p-\bar{x}_p)$, $\Omega\coloneqq\Omega_p\times K\Omega_p\times\{0\}$ is a $[1,H]$ RCI set for the error. \label{lem_RPI_set}
\end{lemma}

The proof can be found in Appendix \ref{app_proof_lem_RPI_set}. Due to lack of space, we do not comment on how to find a suitable $K$ and $\Omega_p$.

\section{Robust Rollout MPC Scheme} \label{sec_MPC_scheme}

Tube MPC relies on the fact that since the difference between real and nominal state is bounded by the $[1,H]$ RCI set $\Omega$ independent of the disturbance realization, the nominal system can be considered in the optimization problem. In order to follow this approach, we tighten the constraints by the $[1,H]$ RCI set $\Omega$ according to
\begin{equation}
\begin{aligned}
\bar{\X} &\coloneqq \X\ominus\Omega = (\X_p\ominus\Omega_p)\times (\U_p\ominus K\Omega_p) \times \I_{[0,b]},\\
\bar{\U} &\coloneqq (\U_p\ominus K\Omega_p) \times \{0,1\},
\end{aligned} \label{tightened_cosntraints}
\end{equation}
where $\bar{\X}$ and $\bar{\U}$ are to be used in the prediction. This ensures that the real state satisfies the original state and input constraints. In addition, we need to make sure that at least every $H$ time steps, the error feedback for the real plant is updated, i.e., that there is a transmission in closed loop at least every $H$ time steps. This is crucial in order to guarantee that the error between real and nominal state remains in the tube, i.e., the $[1,H]$ RCI set.
\begin{remark}
	The exact choice of $H$ poses a tradeoff: for larger $H$, longer inter-transmission intervals can be tolerated, which gives more flexibility for scheduling. However, this would likely increase $\Omega$ as can be seen from $\eqref{inv_ZOH}$, which would result in tighter constraints for the nominal plant. In effect, the feasible set (the set of all states such that the MPC optimization problem is feasible) would shrink as well. \label{rem_size_H}
\end{remark}

To bound the inter-transmission interval in closed loop, we introduce a counter $s(k)$ which keeps track of the time since the last transmission occurred. To be precise, at time $k$, the last transmission was at $k-s(k)-1$. The idea is now to require that the predicted schedules lie in the set $\Gamma_N^H(s)$ as a constraint in the optimization problem. The set $\Gamma_N^H(s)$ is defined as the set of all schedules of length $N$, wherein the first transmission occurs after at most $H-s-1$ steps and in which the remaining ones are at most $H$ steps apart. If $N$ is smaller than $H-s$, the schedule does not have to contain a transmission. Define the ordered sequence of indices for which the schedule $\gamma(\cdot)\in\{0,1\}^N$ is $1$ by
\begin{equation*}
\tau_{\gamma(\cdot)}(\cdot) \coloneqq \{j\in\I_{[0,N-1]}|\gamma(j)=1\} \in\I_{[0,N-1]}^{n_{\gamma(\cdot)}},
\end{equation*}
where $n_{\gamma(\cdot)}\coloneqq\sum_{i=0}^{N-1}\gamma(i)$ is the number of transmissions in $\gamma(\cdot)$. Then, a formal definition of $\Gamma_N^H(s)$ is given by
\begin{align}
&\Gamma_N^H(s) \coloneqq \{\gamma(\cdot)\in\{0,1\}^N|(\tau_{\gamma(\cdot)}(0) \le H-s-1 \nonumber \\
&\land \tau_{\gamma(\cdot)}(j+1)-\tau_{\gamma(\cdot)}(j) \le H, \; \forall j\in\I_{[0,n_{\gamma(\cdot)} - 2]} \label{def_scheduling_constraint} \\
&\land N-\tau_{\gamma(\cdot)}(n_{\gamma(\cdot)}) \le H-1) \lor\; N\le H-s-1\}.  \nonumber
\end{align}
At the end of this section, we will prove that indeed, a transmission is triggered in closed loop at least every $H$ steps with this additional constraint in the optimization problem.

Also in the disturbance-free case, rollout approaches in NCS require special care when designing the underlying MPC law. Since the transmission of an updated control law is not possible at every time step, standard techniques to ensure recursive feasibility and stability cannot be directly applied. In this paper, we follow the approach of MPC with a cyclically time-varying prediction horizon instead of a traditional constant horizon (see \cite{Koegel13,Wildhagen19}). Therein, the prediction horizon at time $k$ is defined by
\begin{equation}
N(k) \coloneqq \overline{N}-k\text{mod}M
\end{equation}
for a cycle length $M\in\I_{\ge 1}$ and a maximum horizon $\overline{N}\in\I_{\ge M}$. We introduce the predicted nominal state and input trajectories at time $k$ as $\bar{x}(\cdot|k)\coloneqq\{\bar{x}(0|k),\ldots,\bar{x}(N(k)|k)\}$ and $\bar{u}(\cdot|k)\coloneqq\{\bar{u}(0|k),\ldots,\bar{u}(N(k)-1|k)\}$.
The objective function is then defined by
\begin{align}
V(\bar{x}(\cdot|k),\bar{u}&(\cdot|k),k) \coloneqq \lambda(\bar{x}(0|k)) \\
&+ \sum_{i=0}^{N(k)-1} \ell(\bar{x}(i|k),\bar{u}(i|k)) + V_f(\bar{x}(N(k)|k)), \nonumber
\end{align}
where $\lambda(\bar{x})\coloneqq\bar{u}_s^\top S \bar{u}_s$, $R\ge S > 0$ is an additional weight on the initial state, and $V_f:\X\to\R$ is the terminal cost.
\begin{remark}
	The term $\lambda(\cdot)$ is included in the objective function to establish convergence later, where the analysis relies on results from economic tube MPC \cite{Bayer14,Dong18_2}. It is indeed essential for this guarantee, but in our case, its effect on the closed loop can be made negligibly small since $S$ may be arbitrarily small as long as it is positive definite.
\end{remark}

Now, we are ready to state the MPC optimization problem. Given the system state $x(k)$, the nominal state $\bar{x}(k)$ and the counter $s(k)$, the problem $\mathcal{P}(x(k),\bar{x}(k),s(k),k)$ solved at each time step $k$ is defined by
\begin{subequations}
	\begin{align}
	&\min_{\bar{x}(\cdot|k),\bar{u}(\cdot|k)} V(\bar{x}(\cdot|k),\bar{u}(\cdot|k),k)  \nonumber \\
	\text{s.t. } &\begin{cases}	x(k)\in\{\bar{x}(0|k)\}\oplus\Omega & \bar{\gamma}(0|k) = 1 \label{constr_IC}\\
												\bar{x}(0|k)=\bar{x}(k) & \bar{\gamma}(0|k) = 0 \end{cases}  \\
	&\bar{x}(i+1|k) = f(\bar{x}(i|k),\bar{u}(i|k)) \label{constr_dynamics} \\
	&\bar{x}(i|k) \in\bar{\X}, \; \bar{u}(i|k) \in \bar{\U}, \quad \forall i\in\I_{[0,N(k)-1]} \label{constr_state_input}  \\
	&\bar{\gamma}(\cdot|k)\in\Gamma_{N(k)}^H(s(k)) \label{constr_scheduling} \\
	&\bar{x}(N(k)|k) \in \bar{\X}_f \label{constr_terminal}
	\end{align}
\end{subequations}
with the closed terminal region $\bar{\X}_f\subseteq\bar{\X}$. We comment on the exact choice of terminal ingredients to ensure recursive feasibility and convergence in the next section.

We denote by the superscript $^*$ the nominal state and input trajectories which solve $\mathcal{P}(x(k),\bar{x}(k),s(k),k)$. The tube MPC then operates with the following scheme.
\begin{algorithm}[h]
	\begin{Algorithm}\label{scheme_MPC}
		\normalfont{\textbf{Robust Rollout MPC Scheme}}
		\begin{enumerate}
			\item[0)] Set $k=0$, enforce $\bar{\gamma}^*(0|0)=1$, choose an arbitrary $\bar{x}(0)$ that fulfills $x(0)\in\{\bar{x}(0)\}\oplus\Omega$ and set $s(0)=0$.
			\item[1)] At time $k$, measure $x(k)$, solve $\mathcal{P}(x(k),\bar{x}(k),s(k),k)$.
			\item[2)] Set $\bar{x}(k+1)=\bar{x}^*(1|k)$. Apply the error feedback
			\begin{equation*}
			u(k) = \begin{cases} [\substack{\bar{u}_c^*(0|k)+K(x_p(k)-\bar{x}_p^*(0|k))\\ 1}] & \bar{\gamma}^*(0|k)=1 \\
					\bar{u}^*(0|k) & \bar{\gamma}^*(0|k)=0 \end{cases}
			\end{equation*}
			to the real system \eqref{system}, i.e., transmit $\bar{u}_c^*(0|k)+K(x_p(k)-\bar{x}_p^*(0|k))$ to the actuator if $\bar{\gamma}^*(0|k)=1$, otherwise do not transmit. Finally, set
			\begin{equation*}
			s(k+1)=\begin{cases} 0 & \bar{\gamma}^*(0|k)=1 \\
							s(k)+1 & \bar{\gamma}^*(0|k)=0 \end{cases}.
			\end{equation*}
			\item[3)] Set $k\leftarrow k + 1$ and go to 1).
		\end{enumerate}
	\end{Algorithm}
\end{algorithm}

\begin{remark}
	We enforce $\bar{\gamma}^*(0|0)=1$ to ensure that an error feedback is transmitted at initial time. This restricts the feasible set at $k=0$ to a subset of $\X_p\times\U_p\times\I_{[c-g,b]}$.
\end{remark}

Note that according to \eqref{constr_IC}, the initial nominal state is an optimization variable only if there is a transmission predicted at initial time, otherwise it is fixed to $\bar{x}(0|k)=\bar{x}(k)=\bar{x}(1|k-1)$. This can be seen as a mixed strategy between \cite[Section 6]{Bayer14}, \cite{Dong18_2,Mayne05} where the initial state is allowed to be an optimization variable at each time instant, and \cite[Section 3]{Bayer14} where the initial state is always fixed. The reason why we employ this strategy is the following: at all time instances where no transmission takes place in closed loop, one cannot reoptimize the nominal plant state, since also the error feedback (which would depend on the newly chosen $\bar{x}_p^*(0|k)$) could not be updated at the actuator due to the lacking transmission. In effect, one could not guarantee anymore that the error between real and nominal state is bounded by $\Omega$. It would be possible in principle to enforce $\bar{x}(0|k)=\bar{x}(k)$ also at times when a transmission does take place. Note, however, that in this case, there would be no feedback whatsoever from the real state back to the controller, which is indeed provided by our strategy.

The following result establishes that under application of Algorithm \ref{scheme_MPC}, there is indeed a transmission at least every $H$ time steps in closed loop.

\begin{lemma}
	Suppose that $\overline{N}\ge H\ge M$ and that for a $\overline{k}\in\I_{\ge 0}$, $\mathcal{P}(x(k),\bar{x}(k),s(k),k)$ is feasible for all $k\in\I_{[0,\overline{k}-1]}$. Then under application of Algorithm \ref{scheme_MPC}, for any $k\in\I_{[0,\overline{k}-H]}$, it holds that
	\begin{equation*}
	\sum_{i=k}^{k+H-1} \gamma(i) \ge 1.
	\end{equation*}
	\label{lem_Q_step_closed_loop}
\end{lemma}
The proof can be found in Appendix \ref{app_proof_lem_closed_loop_Q}.

\begin{remark}
	This result can be described as a transient average constraint known from economic MPC \cite{Mueller14b} or as a weakly-hard real-time constraint from task scheduling  \cite{Bernat01}.
\end{remark}

\section{Recursive Feasibility and Convergence} \label{sec_rec_feas_conv}

In this section, conditions for the terminal region and cost are given, such that recursive feasibility of the MPC optimization problem and convergence can be guaranteed. Furthermore, we comment on satisfaction of the original constraints for system \eqref{system}. We design the terminal ingredients similarly as in \cite{Wildhagen19_2}, where the cycle length $M\coloneqq\lceil \frac{c}{g}\rceil$ is the base period of an a priori chosen periodic transmission schedule, which is feasible according to the token bucket TS.
\begin{assumption}
	There exists a closed set $\bar{\X}_{f,p} \subseteq \X_p\ominus\Omega_p$ containing the origin and a $K_f\in\R^{m_p\times n_p}$, such that it holds that $K_f\bar{\X}_{f,p}\subseteq\U_p\ominus K\Omega_p$, $(A_i+B_i K_f)\bar{\X}_{f,p}\subseteq\X_p\ominus\Omega_p$ for all $i\in\I_{[1,M-1]}$ and $(A_M+B_M K_f)\bar{\X}_{f,p}\subseteq\bar{\X}_{f,p}$.
	\label{ass_term_inv}
\end{assumption}
\begin{assumption}
	There exists a $0<P_f\in\R^{n_p\times n_p}$ such that
		\begin{align}
		&(A_M + B_M K_f)^\top P_f (A_M + B_M K_f) - P_f \\
		&\le -\sum_{i=0}^{M-1} (A_i + B_i K_f)^\top Q (A_i + B_i K_f) - M K_f^\top R K_f. \nonumber
		\end{align}
	\label{ass_term_decr}
\end{assumption}
\begin{remark}
	If the pair $(A_M,B_M)$ is controllable, an admissible choice for $K_f$ and $P_f$ is given in \cite[Lemmas 1,2]{Wildhagen19_2}. If $\X_p\ominus\Omega_p$ and $\U_p\ominus K\Omega_p$ are polytopic, a suitable set $\bar{\X}_{f,p}$ can be constructed using methods as in \cite[Remark 2]{Wildhagen20}. \label{rem_construction terminal_ingredients}
\end{remark}

We choose $V_f(\bar{x})\coloneqq \bar{x}_p^\top P_f \bar{x}_p$ as a terminal cost and
\begin{equation*}
\bar{\X}_f \coloneqq \bar{\X}_{f,p} \times(\U_p\ominus K\Omega_p)\times\I_{[c-g,b]}
\end{equation*}
as a terminal region. We consider the terminal control sequence $\kappa_0(x) \coloneqq[(K_f \bar{x}_p)^\top 1]^\top$ and $\kappa_i(x) \coloneqq [0\; 0]^\top$, for all $j \in \I_{[1,M-1]}$. We note that $K_f$ can be different from $K$.
\begin{remark}
	Note that the terminal region is chosen differently than in \cite{Wildhagen19_2}. This is done to facilitate establishing recursive feasibility of the schedule constraint \eqref{constr_scheduling}. However at the same time, this choice limits our analysis to the notion of convergence instead of asymptotic stability as in \cite{Wildhagen19_2}. This is because \cite[Assumption 5]{Wildhagen19}, which is used in \cite{Wildhagen19_2} to establish stability, is not fulfilled with our choice of terminal region.
\end{remark}

\begin{assumption}
	There exist an $\Omega_p$ and $K$ such that \eqref{inv_ZOH} is fulfilled with a $H\ge M$. Further, $\X_p\ominus\Omega_p$ and $\U_p\ominus K\Omega_p$ are closed and contain the origin. \label{ass_RCI_set}
\end{assumption}

In other words, the admissible inter-transmission interval $H$ with which robust invariance can be guaranteed must be at least as long as the base period $M$ of the token bucket.

\begin{theorem}[Recursive Feasibility and Robust Constraint Satisfaction]
	Suppose that $\mathcal{P}(x(0),\bar{x}(0),0,0)$ is feasible with the additional constraint $\bar{\gamma}^*(0|0)=1$, that Assumptions \ref{ass_term_inv} and \ref{ass_RCI_set} hold, and that $\overline{N}\ge H \ge M$. Then, $\mathcal{P}$ is feasible for all $k\in\I_{\ge 0}$ and the state and input satisfy $x(k)\in\X$ and $u(k)\in\U$ for all $k\in\I_{\ge 0}$. \label{thm_rec_feas}
\end{theorem}
\begin{theorem}[Convergence]
	Suppose, additionally to the conditions of Theorem \ref{thm_rec_feas}, that Assumption \ref{ass_term_decr} holds and that $R\ge S >0$. Then, the nominal state $\bar{x}(k)$ converges to $\mathbb{A}\coloneqq\{0\}\times\{0\}\times\I_{[0,b]}$ as $k\to\infty$. Furthermore, the real state $x(k)$ converges to $\mathbb{A}\oplus\Omega=\Omega_p\times K\Omega_p \times \I_{[0,b]}$ as $k\to\infty$. \label{thm_convergence}
\end{theorem}
The proofs of Theorems \ref{thm_rec_feas} and \ref{thm_convergence} can be found in Appendices \ref{app_proof_thm_rec_feas} and \ref{app_proof_thm_convergence}.

\subsection{Results for window-based TS as in \cite{Antunes14,Gommans17,Koegel19}} \label{subsec_antunes_traffic_spec}

Another TS used in the literature on rollout control is the following: in each disjoint window of length $M\in\I_{\ge 1}$, there are $r\in\I_{[1,M]}$ transmissions possible; these transmissions can be arbitrarily scheduled by the MPC over each interval of length $M$ (cf. \cite{Antunes14,Gommans17,Koegel19}).

We will give the following statements without proof, since they can be easily inferred from the preceding analysis. Unlike in the token bucket TS, the network is not a dynamical component of the NCS and the overall system can be described by the state variables $x\coloneqq[x_p^\top \; u_s^\top]^\top$, such that the right-hand side of \eqref{system} is defined by
\begin{equation*}
f(x,u) \coloneqq \begin{bmatrix} Ax_p + B(1-\gamma) u_s + B\gamma u_c \\ (1-\gamma) u_s + \gamma u_c \end{bmatrix} \text{ and }
w \coloneqq \begin{bmatrix} w_p \\ 0 \end{bmatrix}.
\end{equation*}
The constraint sets are $\X\coloneqq\X_p\times\U_p$, $u(k)\in\U\coloneqq\U_p\times\{0,1\}$ and $w(k)\in\W\coloneqq\W_p\times\{0\}$. A $[1,H]$ RCI set is then $\Omega\coloneqq\Omega_p\times K\Omega_p$ where $\Omega_p$ fulfills \eqref{inv_ZOH}. The tightened state and input constraint sets are defined as in \eqref{tightened_cosntraints}.

For ease of presentation, we assume $r=1$, and note that the following considerations could be easily extended to $r>1$. In the optimization problem $\mathcal{P}$, we allow $S_k$ transmissions in the predicted schedule at time $k$, where $S_k$ is given in \cite[Section IV-B]{Koegel19}, to fulfill the window-based TS in closed loop. We change the terminal region to $\bar{\X}_f\coloneqq\bar{\X}_{f,p}\times(\U_p\ominus K\Omega_p)$, we require that Assumptions \ref{ass_term_inv}-\ref{ass_RCI_set} still hold and set $S=0$. Then Lemma \ref{lem_Q_step_closed_loop} as well as Theorems \ref{thm_rec_feas} and \ref{thm_convergence} (if we define $\mathbb{A}\coloneqq\{0\}\times\{0\}$) hold without further adaptation.

\section{Numerical Example} \label{sec_num_ex}

For a brief numerical analysis, we consider a disturbed double integrator
\begin{equation}
x_p(k+1) = \begin{bmatrix} 1 & 0.1 \\ 0 & 1 \end{bmatrix} x_p(k) + \begin{bmatrix} 0.005 \\ 0.1 \end{bmatrix} u_p(k) + w_p(k)
\end{equation}
subject to the constraints $x_p(k)\in[-8,8]^2$, $u_p(k)\in[-15,15]$ and where the bounded disturbance satisfies $w_p(k)\in[-0.02,0.02]^2$. The parameters of the token bucket TS are $g=1$, $c=3$ and $b=10$, such that the base period is $M=3$. Hence, the maximum inter-transmission interval $H$ for the error feedback must satisfy $H\ge 3$. The numerical results were obtained using Matlab, Yalmip \cite{YALMIP}, SeDuMi \cite{SeDuMi} and MPT3 \cite{MPT3}.

\begin{figure}
	\centering
	\input{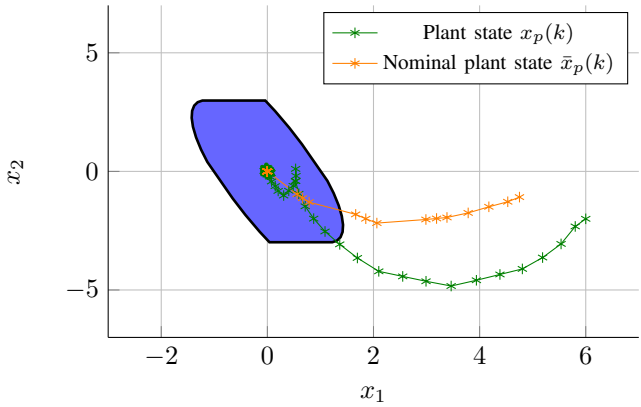}
	\caption{Trajectories of closed-loop system and $\Omega_p$ for $H=5$ (blue).}
	\label{fig_plant_states}
\end{figure}

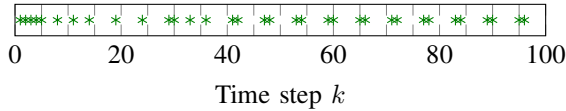
\begin{figure}
	\centering
%
%
\begin{tikzpicture}

\begin{axis}[%
width=\columnwidth,
height=2cm,
xmin=0,
xmax=100,
ytick = \empty,
xtick = {0,5,10,15,20,25,30,35,40,45,50,55,60,65,70,75,80,85,90,95,100},
xticklabels = {0,,,,20,,,,40,,,,60,,,,80,,,,100},
ymin=-1,
ymax=1,
axis background/.style={fill=white},
xminorticks = true,
grid = minor,
xminorgrids = true,
xlabel = {Time step $k$},
]
\addplot [only marks, mark=asterisk, mark options={solid, istgreen}, forget plot]
  table[row sep=crcr]{%
1	0\\
2	0\\
3	0\\
4	0\\
5	0\\
8	0\\
11	0\\
14	0\\
19	0\\
24	0\\
29	0\\
30	0\\
33	0\\
36	0\\
41	0\\
42	0\\
47	0\\
48	0\\
53	0\\
54	0\\
59	0\\
60	0\\
65	0\\
66	0\\
71	0\\
72	0\\
77	0\\
78	0\\
83	0\\
84	0\\
89	0\\
90	0\\
95	0\\
96	0\\
};
\end{axis}
\end{tikzpicture}%
	\caption{Time points of a transmission.}
	\label{fig_transmissions}
\end{figure}

A closed-loop simulation was conducted with a maximum prediction horizon of $\overline{N}=6$, maximum inter-transmission interval of $H=5$, cost matrices $Q=10I$, $R=1$, $S=10^{-6}$, and initial conditions $x_p(0)=[6\;-2]^\top$, $u_s(0)=0$ and $\beta(0)=10$. A suitable terminal region and cost were constructed as discussed in Remark \ref{rem_construction terminal_ingredients}. Figure \ref{fig_plant_states} shows the closed-loop trajectories of the plant, where one can see that the real plant state converges into $\Omega_p$. Actually, the region of ultimate convergence is much smaller than the computed $\Omega_p$, which shows that 
the theoretical statement of Theorem \ref{thm_convergence} can be rather conservative. In Figure \ref{fig_transmissions}, it is indicated at which time points a transmission of control inputs takes place. Due to the schedule constraint, the inter-transmission interval is never longer than $H=5$, which ensures robust constraint satisfaction.

\section{Summary and Outlook} \label{sec_summary_outlook}

In this paper, we have modified tube MPC methods for the control and network access scheduling of disturbed, constrained plants via rollout control. Three steps were necessary to achieve this: First, the definition of a tube (an RCI set), which bounds the error between real and nominal plant despite the fact that the error feedback is zero-order held for up to $H$ time steps; second, an additional schedule constraint in the optimization, which ensures that the inter-transmission interval is upper bounded by $H$ in closed loop; and third, a modified MPC scheme in which the initial state is fixed in case there is no transmission predicted initially.

In a networked control setting, one could also imagine that an actuator is able to implement the error feedback by a prediction of the real plant's state, instead of the pure ZOH error feedback considered in this paper. Furthermore, we have seen that the tubes might grow quite large for a high admissible inter-transmission interval. A different design of the tubes, e.g. as in \cite{Chisci01}, might provide a better estimate of the disturbance's effect on the plant and in effect will come with less conservatism. Finally, we are confident that the case of output feedback in rollout control may also be considered using techniques known from tube MPC.

\bibliographystyle{IEEEtran}
\bibliography{bib_CDC20}

\appendix

\subsection{Proof of Lemma \ref{lem_RPI_set}}
\label{app_proof_lem_RPI_set}

We prove first that if for an arbitrary $i\in\I_{[0,H-1]}$ it holds that $e(i)\in\R^{n_p+m_p}\times\{0\}$, then
	\begin{equation}
	e(i+1)=
	\begin{cases}
	\mathcal{A}_{1}e(0)+w(0) & i=0 \\
	\mathcal{A}_{0}e(i)+w(i) & i\in\I_{[1,H]}
	\end{cases} \label{error_dyn}
	\end{equation}
	where
	\begin{equation*}
	\mathcal{A}_{\bar{\gamma}} \coloneqq \begin{bmatrix}	A+\bar{\gamma}BK & (1-\bar{\gamma})B & 0 \\
	\bar{\gamma}K & (1-\bar{\gamma})I & 0 \\
	0 & 0 & 0	\end{bmatrix}.
	\end{equation*}
	
	Consider first the case $i=0$ for which we have
	\begin{equation*}
		\begin{aligned}
		&e(1) \hspace{-1pt}=\hspace{-1pt} f(x(0),[\substack{\bar{v}_c(0)+K(x_p(0)-\bar{x}_p(0))\\ 1}]\hspace{-1pt}+\hspace{-1pt}w(0)\hspace{-1pt}-\hspace{-1pt}f(\bar{x}(0),[\substack{\bar{v}_c(0)\\ 1}])\\
		&=\begin{bmatrix}	A(x_p(0)-\bar{x}_p(0))+BK(x_p(0)-\bar{x}_p(0)) +w_p(0) \\
		K(x_p(0)-\bar{x}_p(0))  \\
		\min\{\beta(0)+g-\hspace{-1pt}c,b\}-\min\{\bar{\beta}(0)+g- c,b\}	\end{bmatrix}.
		\end{aligned}
		\end{equation*}
		The last line is zero since $\beta(0)-\bar{\beta}(0) = 0$ from $e(0)=x(0)-\bar{x}(0)\in\R^{n_p+m_p}\times\{0\}$.
		
		For $i\in\I_{[1,H-1]}$, we have again $\beta(i)-\bar{\beta}(i) = 0$ and hence
		\begin{equation*}
		\begin{aligned}
		&e(i+1) = f(x(i),[\substack{0\\ 0}])+w(i)-f(\bar{x}(i),[\substack{0\\ 0}])\\
		&=\begin{bmatrix}	A(x_p(i)\hspace{-1pt}-\hspace{-1pt}\bar{x}_p(i))\hspace{-1pt}+B(u_s(i)\hspace{-1pt}-\hspace{-1pt}\bar{u}_s(i))\hspace{-1pt}
		+\hspace{-1pt}w_p(i) \\
		u_s(i)-\bar{u}_s(i)  \\
		0	\end{bmatrix}.
		\end{aligned}
		\end{equation*}
		
		Second, we prove by induction that if $e(0)\in\Omega$, then $e(i)= \mathcal{A}_0^{i-1} \mathcal{A}_1 e(0)+ \sum_{j=0}^{i-1} \mathcal{A}_0^j w(i-j-1)$ and $e(i)\in\Omega$ for all $i\in\I_{[1,H]}$, such that the claim follows immediately. We begin the induction for $i=1$. Denoting $e_p(0)$ the entries of $e(0)$ corresponding to the plant state, we have with $e(0)\in\Omega\subseteq\R^{n_p+m_p}\times\{0\}$ and \eqref{error_dyn}
		\begin{equation*}
		\begin{aligned}
		&e(1) = \mathcal{A}_1 e(0) + w(0) =
		\begin{bmatrix}	(A+BK) e_p(0) \\ K e_p(0) \\ 0	\end{bmatrix} + w(k)	\\
		&\stackrel{\substack{e_p(0)\in\Omega_p,\\ w(0)\in\W}}{\in} (A+BK)\Omega_p \oplus \W_p \times K\Omega_p \times \{0\} \subseteq \Omega,
		\end{aligned}
		\end{equation*}
		where the last inclusion holds due to \eqref{inv_ZOH}.
		
		For the induction step, assume that for $i\in\I_{[2,H]}$, it holds that $e(i-1)=\mathcal{A}_0^{i-2} \mathcal{A}_1 e(0)+ \sum_{j=0}^{i-2} \mathcal{A}_0^j w(i-j-2)$ and $e(i-1)\in\Omega$. Then with \eqref{error_dyn}
		\begin{equation*}
		\begin{aligned}
		&e(i) = \mathcal{A}_0 e(i-1) + w(i-1) \\
		&= \mathcal{A}_0 \bigg(\mathcal{A}_0^{i-2} \mathcal{A}_1 e(0)+ \sum_{j=0}^{i-2} \mathcal{A}_0^j w(i-j-2)\bigg) + w(i-1) \\
		&= \mathcal{A}_0^{i-1} \mathcal{A}_1 e(0)+ \sum_{j=0}^{i-1} \mathcal{A}_0^j w(i-j-1).
		\end{aligned}
		\end{equation*}
		We calculate further, using the definitions of $A_i$ and $B_i$,
		\begin{equation*}
		\begin{aligned}
		&e(i) = \begin{bmatrix}	(A_i+B_i K) e_p(0) \\ K e_p(0) \\ 0	\end{bmatrix} \hspace{-1pt}+\hspace{-1pt} \begin{bmatrix}	\sum_{j=0}^{i-1} A_i w_p(i\hspace{-1pt}-\hspace{-1pt}j\hspace{-1pt}-\hspace{-1pt}1) \\ 0 \\ 0	\end{bmatrix} \\
		&\stackrel{\substack{e_p(0)\in\Omega_p,\\ w_p(i)\in\W_p}}{\in} \hspace{-3pt}(A_i+B_iK)\Omega_p \hspace{-1pt}\oplus\hspace{-1pt} \bigg(\hspace{-1pt}\bigoplus_{j=0}^{i-1} A_j \W_p\hspace{-1pt}\bigg) \hspace{-1pt}\times\hspace{-1pt} K\Omega_p \hspace{-1pt}\times\hspace{-1pt} \{0\} \subseteq \Omega
		\end{aligned}
		\end{equation*}
		where the last inclusion is again from \eqref{inv_ZOH}.

\subsection{Proof of Lemma \ref{lem_Q_step_closed_loop}} \label{app_proof_lem_closed_loop_Q}

Suppose, for contradiction, that for an arbitrary $k\in\I_{[0,\overline{k}-H]}$, $\gamma(k)=1$ and $\gamma(k+i)=0$ for all $i\in\I_{[1,H]}$ under application of Algorithm \ref{scheme_MPC}, i.e., that the inter-transmission interval is longer than $H$ although the predicted schedules always fulfill the constraint \eqref{constr_scheduling}.

In this case, $s(k+i) = i-1$, $i\in\I_{[1,H]}$. The interval until the next time instance at which the full horizon $\overline{N}$ is used, is denoted by $h$, i.e., $N(k+h)=\overline{N}$. According to the cyclic horizon scheme, $h\in\I_{[1,M]}\subseteq \I_{[1,H]}$, where the inclusion is from $H\ge M$. Hence, $\bar{\gamma}^*(l|k+h)=1$ for some $l \in\I_{[0,H-s(k+h)-1]}=\I_{[0,H-h]}$ since $\bar{\gamma}^*(\cdot|k+h)\in\Gamma_{\overline{N}}^H(h-1)$. Note that since $\overline{N}\ge H$, it cannot happen that $N(k+h)=\overline{N}\le H-s(k+h)-1$ due to $s(k+h)=h-1\ge 0$ and in effect, $\bar{\gamma}^*(\cdot|k+h)$ must contain a transmission (cf. \eqref{def_scheduling_constraint}).

Due to the cyclic horizon, $N(k+i) \ge \overline{N}-i+h$ for all $i\in\I_{[h,H]}$. Equality holds if the horizon just decreases in the remaining time while the strict inequality comes into effect if the horizon recovers to $\overline{N}$ at some point. Hence,
\begin{equation*}
N(k+i)\ge \overline{N}-i+h \hspace*{-2pt}\stackrel{\overline{N}\ge H}{\ge}\hspace*{-2pt} H-i+h \hspace*{-2pt}\stackrel{h\ge 1}{\ge}\hspace*{-2pt} H-i+1=H-s(k+i).
\end{equation*}
In consequence, since $\bar{\gamma}^*(\cdot|k+i)\in\Gamma_{N(k+i)}^H(i-1)$, it must hold that $\bar{\gamma}^*(l|k+i)=1$ for some $l\in\I_{[0,H-s(k+i)-1]}=\I_{[0,H-i]}$ for all $i\in\I_{[h,H]}$. Finally, it follows for some $i\in\I_{[h,H]}$ that $\bar{\gamma}^*(0|k+i)=1$ and then according to Algorithm \ref{scheme_MPC}, $\gamma(k+i)=1$. This contradicts the premise, such that the statement of the lemma must hold.

\subsection{Proof of Theorem \ref{thm_rec_feas}} \label{app_proof_thm_rec_feas}

Suppose that $\mathcal{P}$ was feasible at all instances up to time $k\in\I_{\ge 0}$. Consider the candidate initial condition at $k+1$
\begin{equation}
\bar{x}(0|k+1)=\bar{x}^*(1|k). \label{feas_IC}
\end{equation}
It easily holds that $x(k+1)\in\bar{x}^*(1|k)\oplus\Omega$, since $\Omega$ is a $[1,H]$ RCI set from Lemma \ref{lem_RPI_set} and there was a transmission at least every $H$ time steps according to Lemma \ref{lem_Q_step_closed_loop}. In conclusion, the constraint \eqref{constr_IC} is fulfilled by \eqref{feas_IC}.

Now consider the case $(k+1)\text{mod} M\neq 0$ where $N(k+1)=N(k)-1$, for which the candidate input sequence is
\begin{equation}
\bar{u}(\cdot | k + 1)=\{\bar{u}^*(1|k),\ldots,\bar{u}^*(N(k)-1|k)\}, \label{feas_input_1}
\end{equation}
and the case $(k+1)\text{mod} M=0$, where $N(k+1)=\overline{N}=N(k)+M-1$ and for which the candidate input is
\begin{align}
&\bar{u}(\cdot | k + 1)=\{\bar{u}^*(1|k),\ldots,\bar{u}^*(N(k)-1|k), \label{feas_input_2} \\
&\kappa_{0}(\bar{x}(\overline{N}-M-1|k+1)),\ldots,\kappa_{M-1}(\bar{x}(\overline{N}-1|k+1))\}. \nonumber
\end{align}
We refer to the proof of \cite[Theorem 1]{Wildhagen19_2} and \cite{Koegel13} to conclude that $\bar{u}(\cdot | k + 1)$ fulfills \eqref{constr_dynamics}, \eqref{constr_state_input} and \eqref{constr_terminal} in both cases.

Next, we focus on the schedule constraint \eqref{constr_scheduling}. We will show only that the candidate sequences \eqref{feas_input_1} and \eqref{feas_input_2} fulfill the schedule constraint of the ``first'' and ``last'' transmission in \eqref{def_scheduling_constraint}, since it is obvious that the ``middle transmissions'' are just shifted in \eqref{feas_input_1} and \eqref{feas_input_2} compared to $\bar{\gamma}^*(\cdot|k)$, and are still no more than $H$ time steps apart. Consider 1) $N(k)\ge H-s(k)$. Then, $\bar{\gamma}^*(i|k)=1$ for an $i\in\I_{[0,H-s(k)-1]}$ and $\bar{\gamma}^*(j|k)=1$ for a $j\in\I_{[N(k)-H,N(k)-1]}$. In case $\bar{\gamma}^*(0|k)=0$, both candidate input sequences \eqref{feas_input_1} and \eqref{feas_input_2} guarantee that $\bar{\gamma}(i|k+1)=1$ for an $i\in\I_{[0,H-s(k)-2]}=\I_{[0,H-s(k+1)-1]}$ since then, $s(k+1)=s(k)+1$ and $s(k)\ge 0$ and the previously optimal schedule was just shifted in the candidate schedules. In case where $\bar{\gamma}^*(0|k)=1$, we can conclude the same but forgo a derivation due to space constraints. Consider 1a) that  $(k+1)\text{mod}M\neq 0$. With \eqref{feas_input_1}, if $\tau_{\bar{\gamma}^*(\cdot|k)}(n_{\bar{\gamma}^*(\cdot|k)})\ge 1$, then $\bar{\gamma}(j|k+1)=1$ for a $j\in\I_{[N(k)-H-1,N(k)-2]}=\I_{[N(k+1)-H,N(k+1)-1]}$ due to $N(k+1)=N(k)-1$. If $\tau_{\bar{\gamma}^*(\cdot|k)}(n_{\bar{\gamma}^*(\cdot|k)})=0$, one may conclude that $N(k+1)\le H-s(k+1)-1$, such that it must not contain a transmission. In conclusion, $\bar{\gamma}(\cdot|k+1)\in\Gamma_{N(k)-1}^H(s(k+1))$. In case 1b) $(k+1)\text{mod}M=0$, we have $N(k+1)=\overline{N}=N(k)+M-1$ and $\bar{\gamma}(\overline{N}-M|k+1)=1$ due to the scheduled transmission in $\kappa_0$. Since $H\ge M$, we have again $\bar{\gamma}(\cdot|k+1)\in\Gamma_{N(k)+M-1}^H(s(k+1))$. Consider 2) that $N(k)\le H-s(k)-1$, such that $\bar{\gamma}^*(\cdot|k)$ might not contain a transmission at all. If 2a) $(k+1)\text{mod}M\neq 0$, then $N(k+1)=N(k)-1$, such that $N(k+1)=N(k)-1\le H-s(k)-2 \le H-s(k+1)-1$ since $s(k+1)\in\{0,s(k)+1\}$ and $s(k)\ge 0$. In effect, $\bar{\gamma}(\cdot|k+1)$ is not required to contain a transmission and hence, $\bar{\gamma}(\cdot|k+1)\in\Gamma_{N(k)-1}^H(s(k+1))$. If 2b) $(k+1)\text{mod}M = 0$, then $\bar{\gamma}(\overline{N}-M|k+1)=1$ due to $\kappa_0$. It holds on the one hand that $\bar{\gamma}(i|k+1)=1$ for an $i\in\{\overline{N}-M\}=\{N(k)-1\}\stackrel{N(k)\ge 1}{\subseteq}\I_{[0,N(k)-1]}\stackrel{N(k)\le H-s(k)-1}{\subseteq}\I_{[0,H-s(k-2)]}\subseteq\I_{[0,H-s(k+1)-1]}$, where the last inclusion is again from $s(k+1)\in\{0,s(k)+1\}$. On the other hand, $\bar{\gamma}(j|k+1)=1$ for a $j\in\{\overline{N}-M\}\subseteq\I_{[\overline{N}-H,\overline{N}-1]}$ since $H\ge M$. In conclusion, we have $\bar{\gamma}(\cdot|k+1)\in\Gamma_{N(k)+M-1}^H(s(k+1))$ for this last case as well.

So far, we have proven feasibility of $\mathcal{P}(x(k),\bar{x}(k),s(k),k)$ for all $k\in\I_{\ge 0}$. Hence, the second statement of the theorem follows directly from Lemma \ref{lem_Q_step_closed_loop}, the tightened constraint sets \eqref{tightened_cosntraints} and the fact that $\Omega$ is a $[1,H]$ RCI set. 

\subsection{Proof of Theorem \ref{thm_convergence}} \label{app_proof_thm_convergence}

We analyze convergence with the help of so-called rotated cost functions. We refer first to the proof of \cite[Theorem 1]{Wildhagen19_2} to conclude that \cite[Assumption 1]{Wildhagen19} is fulfilled for the nominal system \eqref{nom_system} with the control invariant set $\mathbb{A}$, $\ell_{av}^*=0$ and any storage $\lambda(\bar{x})=\bar{u}_s^\top S \bar{u}_s$ for which $S$ fulfills $R\ge S>0$. It is also proven therein that \cite[Assumptions 2-4]{Wildhagen19} are fulfilled as well. Define $L(\bar{x},\bar{u}) = \ell(\bar{x},\bar{u}) + \lambda(\bar{x})-\lambda(f(\bar{x},\bar{u}))-\ell_{av}^*$ and $\tilde{V}_f(\bar{x})=V_f(\bar{x})+\lambda(\bar{x})$, the rotated objective function
\begin{equation*}
	\tilde{V}(\bar{x}(\cdot|k),\bar{u}(\cdot|k),k) \hspace{-2pt} \coloneqq \hspace{-8pt} \sum_{i=0}^{N(k)-1} \hspace{-8pt} L(\bar{x}(i|k),\bar{u}(i|k)) + \tilde{V}_f(\bar{x}(N(k)|k))
	\end{equation*}
	and the so-called rotated optimization problem $\tilde{\mathcal{P}}(x(k),\bar{x}(k),s(k),k)$:
	\begin{equation*}
	\min_{\bar{x}(\cdot|k),\bar{u}(\cdot|k)} \tilde{V}(\bar{x}(\cdot|k),\bar{u}(\cdot|k),k) \text{ s.t. } \eqref{constr_IC}-\eqref{constr_terminal}.
	\end{equation*}
	
	The crucial observation is now that since $\lambda$ is a storage of the nominal system, $\mathcal{P}$ and $\tilde{\mathcal{P}}$ have the same optimizer (cf. \cite{Dong18_2}). In effect, we may prove convergence of the nominal state to $\mathbb{A}$ using the value function of $\tilde{\mathcal{P}}$, which goes along the same lines as the proof of \cite[Theorem 1]{Wildhagen19}.
	
	The theorem's second statement follows immediately with $x(k)\in\{\bar{x}(k)\}\oplus\Omega$ for all $k\in\I_{\ge 0}$ from Theorem \ref{thm_rec_feas}.

\end{document}